\documentclass[apj,numberedappendix]{emulateapj}
\usepackage{epsfig}
\usepackage{amsmath}
\usepackage{enumerate}
\usepackage{natbib}
\usepackage{hyperref}
\def\bB{{\mathbf B}}
\def\bj{{\mathbf j}}
\def\bE{{\mathbf E}}
\def\rhoGJ{\rho_{\rm GJ}}
\def\bOm{\boldsymbol{\Omega}}
\def\rpc{r_{\rm pc}}

\def\RLC{R_{\rm LC}}
\def\RNS{R_{\rm NS}}
\def\A{{\cal A}}
\def\B{{\cal B}}
\def\Ntrap{N_{\rm trap}}
\def\zz{z^\star}

\def\beq{\begin{equation}}
\def\eeq{\end{equation}}
\def\Eq{Equation}
\def\Eqs{Equations}
\def\Sect{Section}
\def\Sects{Sections}

\newbox\grsign \setbox\grsign=\hbox{$>$} \newdimen\grdimen \grdimen=\ht\grsign
\newbox\simlessbox \newbox\simgreatbox \newbox\simpropbox
\setbox\simgreatbox=\hbox{\raise.5ex\hbox{$>$}\llap
     {\lower.5ex\hbox{$\sim$}}}\ht1=\grdimen\dp1=0pt
\setbox\simlessbox=\hbox{\raise.5ex\hbox{$<$}\llap
     {\lower.5ex\hbox{$\sim$}}}\ht2=\grdimen\dp2=0pt
\setbox\simpropbox=\hbox{\raise.5ex\hbox{$\propto$}\llap
     {\lower.5ex\hbox{$\sim$}}}\ht2=\grdimen\dp2=0pt

\def\simlt{\mathrel{\copy\simlessbox}}

\begin{document}

   \title{Dead zone in the polar-cap accelerator of pulsars}
\author{Alexander Y. Chen, Andrei M. Beloborodov}
\affil{Physics Department and Columbia Astrophysics Laboratory\\
Columbia University, 538  West 120th Street New York, NY 10027}

\begin{abstract}
We study plasma flows above pulsar polar caps using time-dependent 
simulations of plasma particles in the self-consistent electric field. 
The flow behavior is controlled by the dimensionless parameter 
$\alpha=j/c\rhoGJ$ where $j$ is the electric current density and 
$\rhoGJ$ is the Goldreich-Julian charge density. The region of the 
polar cap where $0<\alpha<1$ is a ``dead zone'' --- in this zone particle 
acceleration is inefficient and pair creation is not expected even for 
young, rapidly rotating pulsars. Pulsars with polar caps near
the rotation axis are predicted to have a hollow-cone structure
of radio emission, as the dead zone occupies the central part of the
polar cap. Our results apply to charge-separated flows of electrons ($j<0$) 
or ions ($j>0$). In the latter case, we consider the possibility of 
a mixed flow consisting of different ion species, and observe the 
development of two-stream instability. The dead zone at the polar cap 
is essential for the development of an outer gap near
the null surface $\rhoGJ=0$.
\end{abstract}


\maketitle


\section{Introduction}

Magnetic field lines that pass through the light cylinder of a rotating neutron 
star are twisted and carry electric currents $\bj_B=(c/4\pi)\nabla\times\bB$. 
These currents are sustained by electric field $E_\parallel$ induced along 
the magnetic field $\bB$, and ohmic dissipation $E_\parallel j$ feeds the 
observed pulsar activity. Voltage associated with $E_\parallel$ controls the 
energies of accelerated particles, creation of secondary electron-positron 
pairs, and emission of radio waves. The accelerating voltage has been discussed 
in a number of works on pulsars beginning from early papers in the 1970s 
   (Sturrock 1971; Ruderman \& Sutherland 1975; Arons \& Scharlemann 1979).
 
The key dimensionless parameter of the polar-cap accelerator is
\beq
\label{eq:alpha}
   \alpha=\frac{j_B}{c\rhoGJ}, 
\eeq
where $\rhoGJ=-\bOm\cdot\bB/2\pi c$ is the local corotation charge density of 
the magnetosphere 
   (Goldreich \& Julian 1969). 
For a special value of $\alpha=\alpha_0$ (close to unity) a steady state was 
found for the polar-cap flow with significant particle acceleration 
    (e.g. Arons \& Scharlemann 1979; Muslimov \& Tsygan 1992).
However, $\alpha$ is not, in general, expected to take this special value
    (e.g. Kennel et al. 1979).
Global solutions for approximately force-free pulsar magnetospheres give
$\alpha$ that significantly varies across the polar cap 
   (Timokhin 2006).
In general, $\alpha$ can take any value from $-\infty$ to $+\infty$, depending
on the polar cap distance from the rotation axis and the location inside the
polar-cap region. 

The character of the polar-cap accelerator strongly depends on $\alpha$ 
  (Mestel et al. 1985; Beloborodov 2008, hereafter B08). 
The steady
solution with $\alpha=\alpha_0\approx 1$ is a separatrix between two 
opposite regimes of efficient and inefficient acceleration.\footnote{
      Hereafter we will refer to this separatrix as $\alpha=1$, neglecting the 
      deviation of $\alpha_0$ from unity. The precise $\alpha_0$ 
      is controlled by the curvature of magnetic field lines and the general
      relativistic effects 
          (Muslimov \& Tsygan 1992); 
      its exact value is close to unity and
      is not essential for the rest of the paper.} 
In particular,  if $0<\alpha<1$, $E_\parallel$ is quickly screened in the 
charge-separated plasma flowing from the polar-cap surface. The electric field 
satisfies 
Maxwell equations that read 
  (in the co-rotating frame of the star, see e.g. Fawley et al. 1977; Levinson et al. 2005),
\beq
\label{eq:Gauss}
   \nabla\cdot\bE=4\pi(\rho-\rhoGJ), 
 \eeq
 \beq
 \label{eq:Maxw}
   \frac{\partial\bE}{\partial t}=4\pi(\bj_B-\bj). 
\eeq
If $0<\alpha<1$, there exists a velocity $v=\alpha c$ that allows the charge-separated 
flow $j=\rho v$ to simultaneously satisfy $\rho=\rhoGJ$ and $j=j_B$.
If the flow started from the conducting boundary (which has $E=0$) with 
$v=\alpha c$, no electric field would be generated (then $\nabla\cdot\bE=0$ 
and $\partial\bE/\partial t=0$). The actual boundary has $v\neq \alpha c$, as 
charges are lifted from the polar-cap surface with a small initial $v$,
comparable to the thermal velocity in the surface material.
The deviation of $v$ from $\alpha c$ implies $\rho\neq\rhoGJ$ or 
$j\neq j_B$, which generates electric field. B08 argued that 
\Eqs~(\ref{eq:Gauss}) and (\ref{eq:Maxw}) with $0<\alpha<1$ 
always drive the flow toward $v=\alpha c$, 
like a pendulum is driven by gravity toward its equilibrium position.
The resulting oscillations occur in space or time, 
according to \Eqs~(\ref{eq:Gauss}) or (\ref{eq:Maxw}), respectively.
For example, the steady-state solution for a cold flow exhibits
oscillations in space 
   (Mestel et al. 1985; B08). 
The oscillatory behavior of the flow with $0<\alpha<1$ is, in essence, Langmuir 
oscillations; they are generated near the boundary where the flow is 
injected with $v<\alpha c$ and accelerated toward $v=\alpha c$.

In this paper, we investigate the accelerator with $0<\alpha<1$ in more detail.
In \Sect~2, we write down the steady-state solution for the charge-separated
flow, generalized to non-zero temperature of the polar-cap. We argue that 
the flow is unstable to small perturbations and can develop into a complicated 
time-dependent state with a broad momentum distribution. 
To explore the behavior of the flow, we perform fully kinetic 
time-dependent simulations. The method of simulations is described in \Sect~3, 
and the results are presented in Sections~4 and 5.  
Our simulations confirm the predicted turbulent Langmuir oscillations with a 
small voltage. Particle acceleration in the flow with $0<\alpha<1$
is insufficient to ignite pair creation.
Implications of this ``dead zone'' for radio emission and outer gaps in pulsars
are discussed in \Sect~6.

\section{Steady-state solution for a charge-separated flow}

\subsection{Basic equations}

It is natural first to attempt to construct a simple model 
assuming that the polar-cap flow is steady in the (rotating) frame of the neutron star.
Given the steady magnetic field in this frame, 
and the steady boundary conditions at the stellar surface --- excellent 
static conductor that can supply charges with a given temperature, --- one 
could expect a steady state to be established unless the flow is prone to an instability.

Consider a charge-separated flow from the polar cap that carries electric current 
$j_B$ along magnetic field $\bB$. In a steady state $j=j_B$ (\Eq~\ref{eq:Maxw}).
For simplicity, let us assume that $\bB$ is approximately perpendicular to 
the polar cap and let $z$ measure the altitude above the stellar surface.  
A particle of mass $m$ and charge $e$ that starts with a Lorentz factor 
$\gamma_0\approx 1$ at $z=0$ will accelerate as it moves along the magnetic 
field line,
\beq
\label{eq:a}
    \gamma(z)=\gamma_0+a(z),  \qquad a=-\frac{e(\Phi-\Phi_0)}{mc^2},
\eeq
where $\Phi$ is the electric potential and $E_\parallel=-d\Phi/dz$.
Gravitational acceleration (and centrifugal acceleration in the rotating frame) is 
neglected compared to the electric acceleration. 

The electric potential satisfies Poisson equation,
\beq
\label{eq:Phi}
    \frac{d^2\Phi}{dz^2}=-4\pi(\rho-\rhoGJ),
\eeq
where we assumed that the potential varies along $z$ much faster than it does
in the transverse directions, i.e. the acceleration length $l_\parallel$ is much smaller 
than the characteristic transverse scale of the problem $l_\perp$, which may be 
associated with the size of the polar cap. This condition is satisfied for the flows considered below.\footnote{
      Alternatively, the additional term $\nabla_\perp^2\Phi$ could be moved to the 
      right-hand side of \Eq~(\ref{eq:Phi}) and included in the effective $\rhoGJ$.} 
The term $-\rhoGJ$ may be viewed as a fixed background charge density.
The charge density of the flow itself is given by 
\beq
\label{eq:rho}
     \rho(z)=j_B \int_1^{\infty} \frac{w(\gamma_0)\,d\gamma_0}{v(\gamma_0,z)}.
\eeq
Here $v(\gamma_0,z)$ is the velocity of particles that start at $z=0$
with initial Lorentz factor $\gamma_0$; note that $v^2/c^2=1-\gamma^{-2}$ where 
$\gamma(z)$ is given by \Eq~(\ref{eq:a}).
Function $w(\gamma_0)$ describes the probability distribution of $\gamma_0$. 
The width of this distribution is controlled by the temperature of polar cap $T$. 
For example, $w=\delta(\gamma_0-1)$ describes a cold polar cap ($T=0$) where 
all particles have $\gamma_0=1$. 

We multiply both sides of \Eq~(\ref{eq:Phi}) by $da/dz=-(e/mc^2)d\Phi/dz$,
substitute \Eq~(\ref{eq:rho}), and find
\beq
\label{eq:diff}
     \frac{mc^2}{2e}\,\frac{d}{dz}\left(\frac{da}{dz}\right)^2
   =4\pi\left[
   \frac{j_B}{c}\int_1^{\infty} \frac{dp}{dz}(\gamma_0,z)\,w(\gamma_0)\,d\gamma_0
         -\frac{da}{dz}\,\rhoGJ\right].
\eeq
On the right-hand side, we used $da/dz=-d\gamma/dz$ (\Eq~\ref{eq:a}) and 
$d\gamma/v=dp/c$. Integration of \Eq~(\ref{eq:diff}) in $z$ gives
\beq
\label{eq:steady}
    \frac{\lambda_p^2}{2}\left(\frac{da}{dz}\right)^2    
         =\int_1^{\infty} \left[p(\gamma_0,z)-p_0\right]\,w(\gamma_0)\,d\gamma_0
           -\frac{a(z)}{\alpha},
\eeq
where 
\beq
   p^2(\gamma_0,z)=\gamma^2-1=\left[\gamma_0+a(z)\right]^2-1.
\eeq
In \Eq~(\ref{eq:steady}) we used $a(0)=0$ and the boundary condition $da/dz(0)=0$ 
(the stellar surface is modeled as a perfect conductor that can freely emit charges 
with $E_\parallel(0)=0$). We also used $j_B(z)\approx const$ and 
$\rhoGJ(z)\approx const$, as $j_B$ and $\rhoGJ$ do not significantly vary
on the characteristic acceleration length $\lambda_p$, which is defined by
\beq
\label{eq:lambda}
   \lambda_p^2=\frac{mc^3}{4\pi e j_B}.
\eeq
This length may be thought of as the plasma skin depth; it is related to the
plasma frequency $\omega_p$,
\beq
   \lambda_p=\frac{c}{\omega_p}, \qquad \omega_p^2=\frac{4\pi n e^2}{m},
\eeq
where $n=j_B/ec$ is the characteristic plasma density.

A quick estimate for $j_B$ and $\lambda_p$ in pulsars may be obtained from
the following consideration. The magnetic flux through the polar cap $\Psi$ equals 
the flux through the light cylinder $\RLC=c/\Omega$. 
The bundle of open field lines is strongly twisted at the light cylinder
(toroidal component comparable to poloidal), and hence it carries 
electric current $I\sim c\Psi/2\pi\RLC$, according to Stokes theorem.
The current density near the star satisfies $j_B/B\approx I/\Psi$ 
(which follows from the fact that $\bj$ flows along $\bB$); this yields
\beq
\label{eq:j_B}
     j_B\sim \frac{\Omega B}{2\pi}.
\eeq
Then the plasma skin depth in the polar-cap accelerator may be expressed as 
\begin{equation}
    \lambda_p \sim \frac{c}{(\Omega\omega_B)^{1/2}}, 
    \qquad \omega_B=\frac{eB}{mc}.
\end{equation}
The scale $\lambda_p$ is much smaller than the typical size of the polar cap 
$\rpc\sim(\RNS^3\Omega/c)^{1/2}$, where $\RNS\sim 10^6$~cm is the radius of 
the neutron star.

\begin{figure}[h]
\epsscale{1.15}
\begin{center}
\plotone{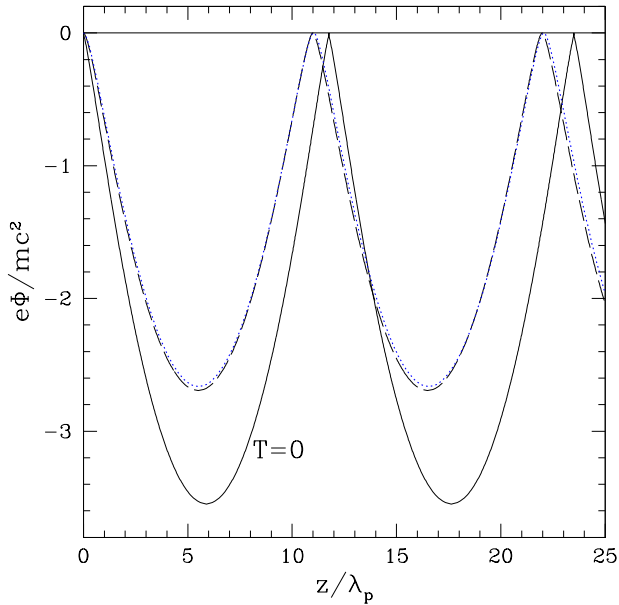}
\caption{Steady-state solution for the charge-separated polar-cap flow with 
$\alpha=0.8$. Two cases are shown: cold polar cap $T=0$ (solid curve)
and hot polar cap $kT/mc^2=0.03$ (dashed curve), which corresponds
to average injection momentum $0.22mc$. 
Dotted curve shows the solution for a flow where all particles are injected with the same $p_0=0.22$.
}
\label{fig:steady}
\end{center}
\end{figure}

\subsection{Cold and warm solutions}\label{sec:solutions}

Once the injection distribution $w(\gamma_0)$ is specified,
it is straightforward to numerically integrate \Eq~(\ref{eq:steady}) and find $a(z)$.
In our sample models we chose 
$w(\gamma_0)=(kT)^{-1}\exp[-(\gamma_0-1)/kT]$ with $kT/mc^2=0$ (cold) and 
0.03 (warm); the average injection momentum $p_0$ in the warm model equals 
$0.22mc$. Figure~\ref{fig:steady} shows $\Phi(z)$ for the cold and warm solutions.

Figure~\ref{fig:steady} also shows a third model where all particles injected at the polar cap 
have $p_0=0.22$, i.e. $w(\gamma_0)$ is a delta-function. 
In this model, \Eq~(\ref{eq:steady}) simplifies to 
\beq
\label{eq:cold}
    \frac{\lambda_p^2}{2}\left(\frac{da}{dz}\right)^2 = p(\gamma_0,z)-p_0
                                                                               + \frac{a(z)}{\alpha},
\eeq
[same as \Eq~(3) in B08]. This flow is everywhere cold, i.e. its momentum 
distribution is described by $f(p^\prime)=n\,\delta[p^\prime-p(z)]$. As one can 
see in Figure~\ref{fig:steady}, the cold model with $p_0\neq 0$ provides an 
excellent approximation to the exact warm model that has the same {\it average}
value of $p_0$. 

The cold flow solution was discussed in earlier works 
   (Mestel et al. 1985; B08).
For $1-\alpha\ll 1$, the oscillation period is approximately given by (B08)
\beq
    z_0 \approx 2^{3/2}\frac{\lambda_p}{1-\alpha}, \qquad 1-\alpha\ll 1.
\eeq
The precise period is obtained by numerical integration; e.g. 
$z_0=11.0\,\lambda_p$ for $\alpha=0.8$. 
The momentum of the steady cold flow $p(z)$ oscillates between the injection 
momentum $p_0\ll 1$ and a maximum value $p_{\max}$. The minima and 
maxima are where $da/dz=0$, and from \Eq~(\ref{eq:cold}) one finds
\beq
\label{eq:pmax}
    p_{\max}=\frac{2\alpha\,\gamma_0-(1+\alpha^2)p_0}{1-\alpha^2}.
\eeq

The above equations assumed $\alpha(z)=const$. In real pulsars, $\alpha$ 
varies due the field-line curvature and general relativistic effects 
(Muslimov \& Tsygan 1992). The length-scale of this variation $(d\alpha/dz)^{-1}$ 
is typically larger or comparable to the stellar radius, which exceeds $\lambda_p$ 
by several orders of magnitude.
When $\alpha$ varies with $z$, the analytical integration of the dynamic equation 
is not possible and one has to solve the two coupled differential equations (B08),
\begin{eqnarray}
\label{eq:dpdz}
    \frac{dp}{dz} & = & \frac{\sqrt{1+p^2}}{pc}\frac{eE_\parallel}{mc},  \\
\label{eq:dEdz}
    \frac{dE_\parallel}{dz} & = & \frac{4\pi j_B}{c}\left(\frac{\sqrt{1+p^2}}{p} 
                                                - \frac{1}{\alpha(z)}\right)
\end{eqnarray}
The solution is similar to the case where $\alpha$ is constant, as long as $0<\alpha<1$.
   The momentum $p(z)$ quasi-periodically passes through maxima and minima. 
The only difference is that the period $z_0$ and $p_\mathrm{max}$ now gradually
change with $z$ (see Figure~1 in B08).


\subsection{Stability of the flow}

Although the cold flow solution with $p_0=0.22$ reproduces very well the electric 
potential $\Phi(z)$ of the exact warm solution with the same average $p_0$,
the warm and cold flows are qualitatively different. Their different
momentum distribution functions $f(p,z)$ leads to
a qualitatively different response to small perturbations.
 
Consider first the cold-flow solution
shown by the blue dotted curve in Figure~\ref{fig:steady}.
Since all particles are injected with the same momentum $p_0=0.22$, 
all of them follow a single trajectory in the phase space $(z,p)$.
They periodically reach the minimum momentum $p_0$ at 
$\zz_k=kz_0$ ($k=0,1,...$) where potential $\Phi$ reaches maximum.
There are no particles with momenta $p\approx 0$, so 
a small perturbation cannot force any particles to reverse their direction of motion,
and hence the perturbation will be advected along the flow. 
This flow is expected to be stable.

In contrast, the warm flow (dashed curve in Figure~\ref{fig:steady}) has a broad 
distribution of $p_0$ that extends from $p_0=0$. At each peak of the electric 
potential (at $z=\zz_k$)
there is a population of particles with nearly zero velocities. Consider a 
perturbation at $z\approx \zz_k$. For example, suppose 
a small bunch $\A$ of particles with momenta in a range $(p_1,p_1+\Delta p)$ 
are slightly pushed forward while the rest of particles are unperturbed.
This perturbation implies a local increase in electric current $j>j_B$ and hence
$\partial E_\parallel/\partial t<0$ (\Eq~\ref{eq:Maxw}), generating negative
electric field $\delta E_\parallel$ at $z\approx \zz_k$ that tends to restore the 
condition $j=j_B$. In contrast to the initial perturbation,
the induced $\delta E_\parallel$ affects {\it all} local particles, 
regardless of their momenta, not just bunch $\A$. This has two implications: 
(1) The induced $E_\parallel<0$ will easily and quickly reduce $j$ back to $j_B$ but 
will be unable to decelerate bunch $\A$ to the momentum it would have in the 
steady state flow --- bunch $\A$ will continue to move to $z>\zz_k$ with a larger 
momentum. 
(2) $\delta E_\parallel<0$ will give very slow particles $p\approx 0$ negative velocities,
creating a new bunch $\B$ that slides backward down the potential hill. Bunch $\B$
creates $j<j_B$ at $z<\zz_k$, and the system reacts there by inducing a small 
$\delta E_\parallel > 0$, which accelerates all local particles, regardless their momenta,
not just bunch $\B$. As a result, $j$ quickly recovers to $j_B$, however, bunch $\B$
is not stopped from moving backward and away from $z=\zz_k$. 

One concludes that the perturbation creates a permanent damage 
to the steady state that broadens the momentum distribution by creating backflowing 
particles. 
This perturbation is {\it not} advected away along the flow, and can develop further. 
The backflowing particles turn out to be trapped between two peaks of the 
electrostatic potential. Further development can be studied with kinetic time-dependent 
simulations; it eventually completely destroys the steady state solution.


\section{Numerical setup}\label{sec:setup}

Our numerical method is similar to that used by 
   Beloborodov \& Thompson (2007, hereafter BT07). 
The plasma is modeled as a large number $N\sim 10^6$ of individual 
particles that flow along the magnetic field lines.
We assume that the magnetic field is fixed in 
the co-rotating frame of the star; thus $j_B$ and $\rhoGJ$ are fixed. Then the 
problem becomes essentially one-dimensional, as discussed in detail in BT07.
In the present paper, we consider only charge-separated flows, with no pair creation. 
Three other differences from the magnetar simulation in BT07 are as follows:
(1) The magnetar problem had $\alpha\gg 1$
($\rhoGJ$ was negligible compared with $j_B/c$); 
in contrast, $\rhoGJ$ is crucial for polar-cap flows considered here.
(2) The presence of gravity was essential for the closed-field circuit considered 
in BT07, where the global plasma flow was 
studied on a scale comparable to the radius of the star;
in the problem considered here the electric fields are 
screened on a much smaller scale $\sim \lambda_p$ and the gravitational 
acceleration plays no role.
(3) The flow behavior on the small scales $z\ll \RNS$ may be studied using 
a smal computational box $H\ll \RNS$ with an open outer boundary (see below).

In the absence of pair creation, the flow is composed of particles lifted from 
the surface. In most simulations presented below we assume that all particles 
have the same mass $m$ and charge $e$.
The particle motion is described by the equation,
\begin{equation}
\label{eq:p}
    \frac{dp_i}{dt} = \frac{eE_\parallel(z_i)}{mc},   \qquad i=1,..,N,
\end{equation}
where $p_i$ is the momentum of the $i$-th particle in units of $mc$, and 
$E_\parallel(z_i)$ is the self-consistent electric field at the particle location $z_i$.
The field is found by integrating Gauss law (\Eq~\ref{eq:Gauss}) 
along the magnetic field line,
\begin{equation}
 \label{eq:E}
   E_\parallel(z_i) = 4\pi \left[ eN(z_i) - \rhoGJ z_i \right]. 
\end{equation}
Here $N(z_i)$ is the column density of particles between $z=0$ and $z=z_i$, 
and we used the boundary condition $E_\parallel(0)=0$, as the material below the 
stellar surface is assumed to be a very good conductor that can emit free charges. 
Divergence of the perpendicular component of electric field $E_\perp$
is neglected in \Eq~(\ref{eq:E}) (see BT07 for discussion of this approximation).
The approximation $|\nabla_\perp\cdot\bE_\perp|\ll |dE_\parallel/dz|$ is valid if the 
characteristic scale of the flow acceleration $z_0$ is smaller than the transverse
scale $l_\perp$, which is limited by the polar-cap size $\rpc$; 
the condition $z_0\ll \rpc$ is satisfied in the dead-zone models presented below. 
We also assume that $\rhoGJ$ is approximately constant on scale $z_0$.
\Eqs~(\ref{eq:p}) and (\ref{eq:E}) in essence describe a relativistic, time-dependent
diode problem with an additional fixed background charge density $-\rhoGJ$. 

As we track the motion of all particles individually, the continuity equation is 
automatically satisfied; for a charge-separated flow it is equivalent to charge 
conservation,
\beq
\label{eq:charge}
    \frac{\partial\rho}{\partial t}+\frac{\partial j}{\partial z}=0.
\eeq 
\Eq~(\ref{eq:Maxw}) follows from \Eqs~(\ref{eq:Gauss}) and (\ref{eq:charge}), 
so we will not need \Eq~(\ref{eq:Maxw}).
Instead, the parameter $j_B$ enters the problem as a boundary condition. 
The magnetic field lines are frozen in the excellent conductor below the stellar 
surface, which sustains $j(0)=j_B$.
This condition is enforced in the simulation by injecting the charges in 
the computational box at $z=0$ with the fixed rate $j_B$ (BT07). 

The electric current $j_B$ is enforced at one boundary $z=0$. Since
the computational box has a finite size $H$, we also have to choose a
boundary condition at $z=H$ and the value of $H$. In all sample models 
shown in this paper we use the simplest boundary condition:
particles moving out of the box are lost and no particles enter the box at 
$z=H$. This condition may be refined by allowing a small inflow of returning 
particles at the outer boundary. We ran test simulations that show that the 
refinements are not important as long as the boundary is sufficiently far, 
so that $H$ is much larger than the characteristic scale of the flow acceleration. 

In the one-dimensional model, the transverse
gradients are neglected and the flow effectively has a slab geometry. 
Then it is sufficient to follow particles flowing through a small area $A$ of the slab.
This allows one to chose a reasonable number of particles in the computational 
box, $N\sim AHn$, e.g. $N\sim 10^6$, so that their dynamics can be followed 
in a reasonable computational time. On the other hand, 
$N$ should be large enough so that the plasma scale $\lambda_p$ contains 
many particles $N_p=A\lambda_p n$. 

In summary, we choose $N$ and $H$ so that 
\beq
    \frac{H}{\lambda_p}\gg 1, \qquad N_p=\frac{\lambda_p}{H}\,N \gg 1.
\eeq  
In this limit, the results are expected to be independent of the choice of $N$ and 
$H$ (we verified this by varying the two parameters). For most of our simulations 
$H=100\lambda_p$ and $N\sim 10^6$. Another requirement is a small time step 
of the simulation, $\Delta t\ll\omega_p^{-1}$, so that plasma oscillations are well 
resolved.


\section{Results}

\subsection{Steady state and stability tests}

In our simulations and in reality the plasma above pulsar polar caps is collisionless. 
In the absence of pair creation it must satisfy the Vlasov equation,
\begin{equation}
\label{eq:Vlasov}
    \frac{\partial F}{\partial t} + \mathbf{v}\cdot \nabla F + \frac{d\mathbf{p}}{dt}\cdot \nabla_\mathbf{p} F = 0,
\end{equation}
where $F(t,z,p)$ is the particle distribution function in phase space. The 
electric current is $j(t,z)=\rho \bar{v}$ where $\bar{v}(t,z)$  is the average 
velocity of the particles. As a first simple test, consider a uniform flow 
with $\rho(z)=\rhoGJ$, $\bar{v}(z)=\alpha c$, and $E_\parallel(z)=0$. 
It is easy to see from \Eqs~(\ref{eq:p}), (\ref{eq:E}) and (\ref{eq:Vlasov}) that 
the flow must remain in this state. This behavior is reproduced by our simulations. 
The steady uniform flow can have any momentum distribution $F(p)$
as long as $\bar{v}=\alpha c$. Note that it requires a continual injection of particles
at $z=0$ with the average velocity $\bar{v}=\alpha c$ (which also requires $0<\alpha<1$).

\begin{figure}
\epsscale{1.03}
\begin{center}
   \plotone{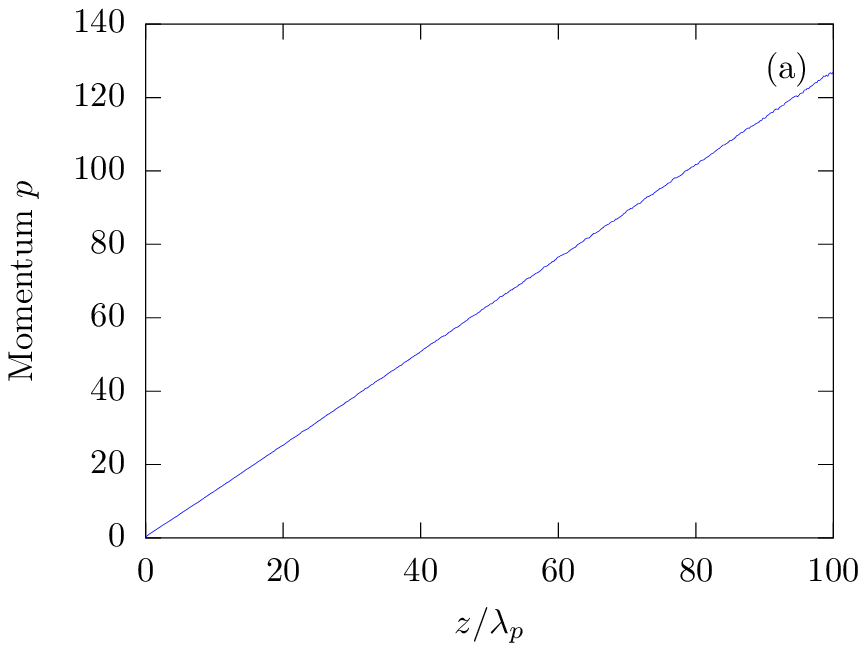}\\
   \plotone{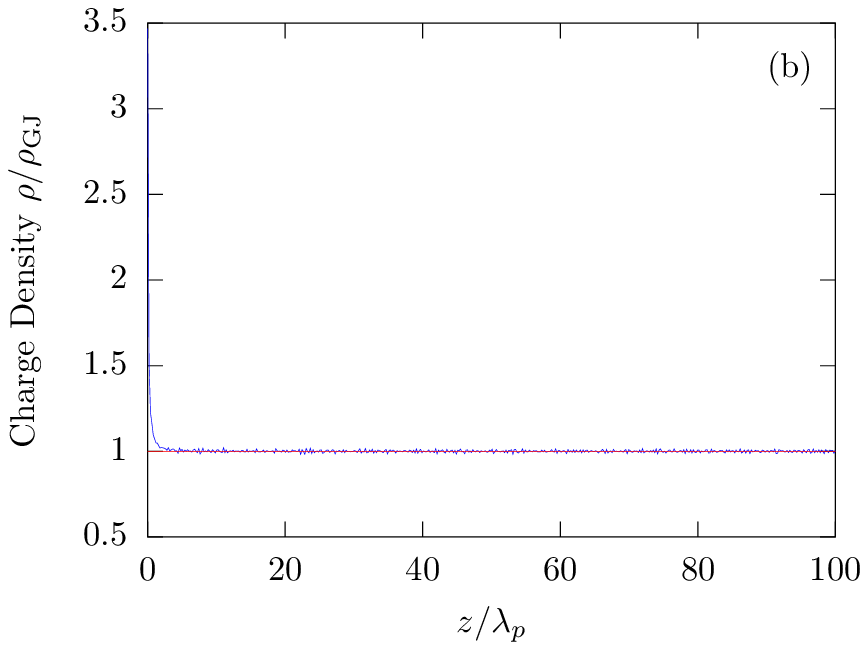}
\caption{Test run for a cold-flow model with $\alpha=1$ and $v_0=c/6$.
The flow relaxed to a steady state in the entire box $H=10^2\lambda_p$ 
on the light-crossing timescale, $H/c$; the state of the system is 
shown at $t = 10H/c$.
(a) Flow momentum per particle $p(z)$ in units of $mc$. 
(b) Charge density $\rho(z)$ in units of $\rhoGJ$.}
\label{fig:alpha1}
\end{center}
\end{figure}

As a second test, consider a ``cold'' flow where all particles move with 
momentum $p(z)$, with zero momentum dispersion. Suppose the flow is injected 
at $z=0$ with velocity $v_0<\alpha c$. Then $E_\parallel$ must be generated,
accelerating the flow. In a steady state, the solution for the cold flow must have 
the form, $F(z,p^\prime)=n(z)\,\delta[p^\prime-p(z)]$, where $p(z)$ and $n(z)$ can 
be described analytically. We first test the special case $\alpha=1$
   (Michel 1974).
The flow is accelerated by the self-consistent $E_\parallel(z)$, and $p$ exceeds 
unity at $z\sim\lambda_p$. At heights $z\gg \lambda_p$, velocity approaches $c$, 
charge density of the flow $\rho=j_B/v$ approaches $\rhoGJ$, and 
electric field $E_\parallel$ asymptotes to a constant value, 
\beq
   E_\parallel=\left[\frac{8\pi m cj_B}{e}\left(\gamma_0-p_0\right)\right]^{1/2}
                      \left[1+{\cal O}(p^{-1})\right],
\eeq 
where $\gamma_0=(1-v_0^2/c^2)^{-1/2}$ and $p_0=\gamma_0\beta_0$.
Then the flow momentum keeps growing linearly with $z$,
\beq
     p(z)=\left[2(\gamma_0-p_0)\right]^{1/2}\,\frac{z}{\lambda_p}, \qquad z\gg\lambda_p.
\eeq
This analytical
solution is reproduced by our simulations. After an initial relaxation period 
(comparable to the light crossing time of the computational box) the system 
forgot initial conditions and relaxed to the steady state shown in Figure~\ref{fig:alpha1}
(in this example, $v_0=1/6$).
The charge density of the flow is large near the polar cap surface and asymptotes
to $\rhoGJ$ at $z\gg\lambda_p$, as expected.

\medskip

\begin{figure}
\begin{center}
\epsscale{1.1}
 \plotone{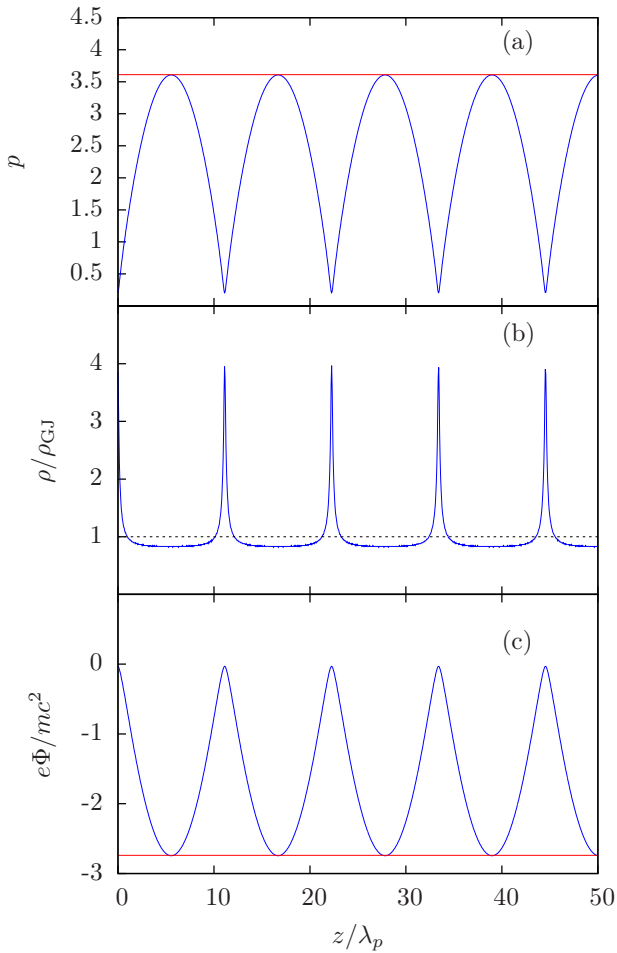}
\caption{Cold flow with $\alpha=0.8$ and $\beta_0=0.2$ at time $t=1.45H/c$
(a) Momentum $p$ (in units of $mc$). 
Red line shows the maximum value predicted by \Eq~(\ref{eq:pmax}).
(b) Charge density. (c) Electrostatic potential.
Red line shows the minimum value predicted by the analytical model of \Sect~2.
}
\label{fig:alpha0.8cold}
\end{center}
\end{figure}

Then we studied cold flows with $0<\alpha<1$ with a
fixed injection velocity $\beta_0$. 
We chose in our sample numerical model $\alpha=0.8$ and $\beta_0=0.2$.
The computational box was initially empty; the plasma injected at $z=0$ 
filled the box on the dynamical timescale 
$\sim H/c$ and established a steady state shown in Figure~\ref{fig:alpha0.8cold}.
The steady state is in perfect agreement with the analytical model of \Sect~2.
The charge density $\rho(z)$ has spikes at $z=kz_0$ ($k=0,1,...$) where 
the flow has the minimum velocity $\beta_0$; the height of each spike is 
$\rho_{\max}=j_B/\beta_0=(\alpha/\beta_0)\rhoGJ$.
The charge spikes are associated with maxima of the electric potential (Figure~\ref{fig:alpha0.8cold}c). 
The oscillating momentum has maxima $p_{\max}=3.6$, in excellent agreement 
with \Eq~(\ref{eq:pmax}). The period of oscillation is $z_0\approx 11\lambda_p$, same
as found using the method of \Sect~2.

As anticipated in \Sect~2.3, we find that the steady state becomes unstable if we 
reduce $\beta_0$ to zero. Then any small perturbation (e.g. due to numerical error) 
completely destroys the steady state; instead, a time-dependent state forms, with a 
broadened momentum distribution function. 
A steady flow with a finite $\beta_0\neq 0$ can also be destroyed, although in this 
case a finite, 
sufficiently large perturbation is required. In fact, this case provides a better
setup for a numerical analysis of the instability, 
as we can control the form of the initial perturbation and then observe how it destroys 
the flow that was stable before the perturbation was applied. 
We made such an experiment with the flow with $\alpha=0.8$ and $\beta_0=0.2$. 
We applied a perturbation that was localized in space and time --- a small ``kick'' 
$\delta p$ was given to all particles located in a small region $\delta z=\lambda_p/2$; 
in this experiment $\delta p$ had a Gaussian distribution with the mean value and 
dispersion equal to 0.02. 
We observed the following evolution. As the localized 
perturbation moved along with the background flow, it was greatly amplified when it 
reached the potential maximum (which corresponds to the minimum $p_0\approx 0.2$ 
of the steady-state solution, see Figure~\ref{fig:alpha0.8cold}), and some 
particles acquired a negative momentum, i.e. reversed their direction of motion. 
Most of the reversed particles became trapped between two potential maxima, 
and some of them were able to penetrate even further back, beyond the 
preceding potential peak. The perturbation further spread in the phase space
and the damage to the initial steady-state solution was further amplified with time, 
in particular near the potential maxima.
Eventually, the entire flow became strongly time-dependent
and the regular periodic structure of potential peaks disappeared.

The amplification of small (linear) perturbations at the potential maximum can be 
understood as follows.
Consider a particle whose Lorentz factor differs from that of the background cold 
flow by a small $\delta\gamma$. As the particle moves along with the flow, its deviation
$\delta\gamma$ remains constant, because it travels in the same electrostatic 
potential of the background flow (cf. \Eq~\ref{eq:a}). 
Using the relation $d\gamma/dp=\beta$, we 
find the perturbation of momentum $\delta p$ that corresponds to $\delta\gamma$,
\beq
    \delta p = \frac{\delta\gamma}{\beta}\propto \beta^{-1}.
\eeq 
It grows as the particle (and the background flow) decelerates near the potential 
maximum; the corresponding amplification factor $\beta_0^{-1}$ is particularly 
large if $\beta_0$ is small.

The generation of backflowing particles at the potential peaks $\zz_k$ plays a key 
role in disrupting the steady state. We also observe that, in 
a flow with a finite minimum velocity $\beta_0>0$, 
only a finite, sufficiently large perturbation can destroy the steady state. 
The perturbation would need to steal from particles energy 
$\gamma_0-1\approx \beta_0^2/2$ so that they can be reflected 
by the potential hill. The energy gap $\gamma_0-1$ stabilizes the flow against 
infinitesimal perturbations, and only a sufficiently strong kick disrupts the flow.  

The trapped/backflowing particles have a deteriorating effect on the steady state 
because they are not advected away with the flow and instead repeatedly approach 
the same potential peaks, amplifying the perturbations. In addition, one can view 
the trapped particles as extra
charge that distorts the electric field. Let $\Ntrap$ be the number of particles
trapped between two potential peaks $\zz_{k-1}$ and $\zz_k$; they create electric 
field $E^\prime=4\pi e\Ntrap$ at $z>\zz_k$. The corresponding distortion of the 
electrostatic potential $\Phi^\prime=-E^\prime z$ grows linearly with $z$ and 
becomes significant at sufficiently large $z$ even if $\Ntrap$ is small.
The distance $z$ required to produce 
$e\Phi^\prime \sim mc^2$ is $z\sim (N_p/\Ntrap)\lambda_p$. 
This behavior is qualitatively confirmed by our numerical experiments with 
larger simulation boxes $H$ --- the flow 
was found to become more unstable with increasing $H$.

\subsection{Time-dependent state with warm particle injection}\label{sec:warm}

In a more realistic model, particles are lifted from the polar cap with a thermal 
velocity dispersion $\Delta v_0\sim v_0$. The flow still starts with a small velocity 
$\bar{v}\ll c$ and hence with a large charge density 
$\rho\gg\rhoGJ$, which self-consistently generates the accelerating electric field.
The basic acceleration mechanism is the same as for the cold flow shown in 
Figures~\ref{fig:alpha1} and \ref{fig:alpha0.8cold}. However, there is a new feature: 
particles with different 
initial velocities behave differently in the collective electric potential, and the charge 
density $\rho(z)$ is changed from the cold-flow solution, even though $\Delta v_0\ll c$. 
Some particles have $v\approx 0$ and can reverse their motion in the regions of 
growing potential ($E_\parallel<0$), which greatly complicates the behavior of the 
distribution function $F(z,p)$.

In our simulations, we modeled the warm injection by a one-dimensional Maxwell 
distribution, which is a simple Gaussian with dispersion $\Delta v_0$ equal to the 
mean value $\bar{v}_0$; we chose $\bar{v}_0=0.2c$.
As initial conditions we took the steady-state solution (\Sect~2).
The main parameter of the flow is $\alpha$, and we performed simulations
for several values of $\alpha$ in the range $0<\alpha<1$. 

\begin{figure}
\begin{center}
\epsscale{1.15}
\plotone{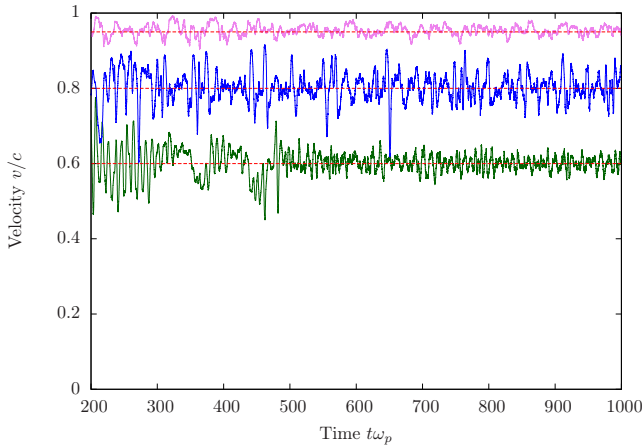}\\
 \caption{Evolution of the hydrodynamical velocity $\bar{v}$ of the flow measured in the middle of the computational box. Three models are shown:  $\alpha = 0.95$ (purple), $0.8$ (blue) and $0.6$ (dark green). In all three cases, the time-average value of $\bar{v}$ equals $\alpha$ (red horizontal lines).
}
\label{fig:meanvelocity}
\end{center}
\end{figure}

As expected, the steady state was quickly destroyed and the flow kept oscillating 
in space and time. The basic parameters of the flow 
remained, however, similar to the steady cold model. The average charge density 
(averaged over oscillations) is nearly equal to $\rhoGJ$ and the average velocity 
$\bar{v}$ is nearly equal to $\alpha c$, so that the condition $\bar{j}=j_B$ is satisfied.
Figure~\ref{fig:meanvelocity} shows the evolution of the hydrodynamic velocity 
$\bar{v}(t)$ 
measured at a fixed location $z_1$ (we chose $z_1=50\lambda_p$, in the middle 
of the computational box; $\bar{v}$ was calculated by averaging over 
particles inside a small bin around $z_1$, of width $2\lambda_p$). 
The hydrodynamic velocity $\bar{v}(t)$ oscillates around $\alpha c$;
these oscillations have a relatively small amplitude $\delta v\ll \bar{v}$.

\begin{figure}
\begin{center}
\epsscale{1.15}
  \plotone{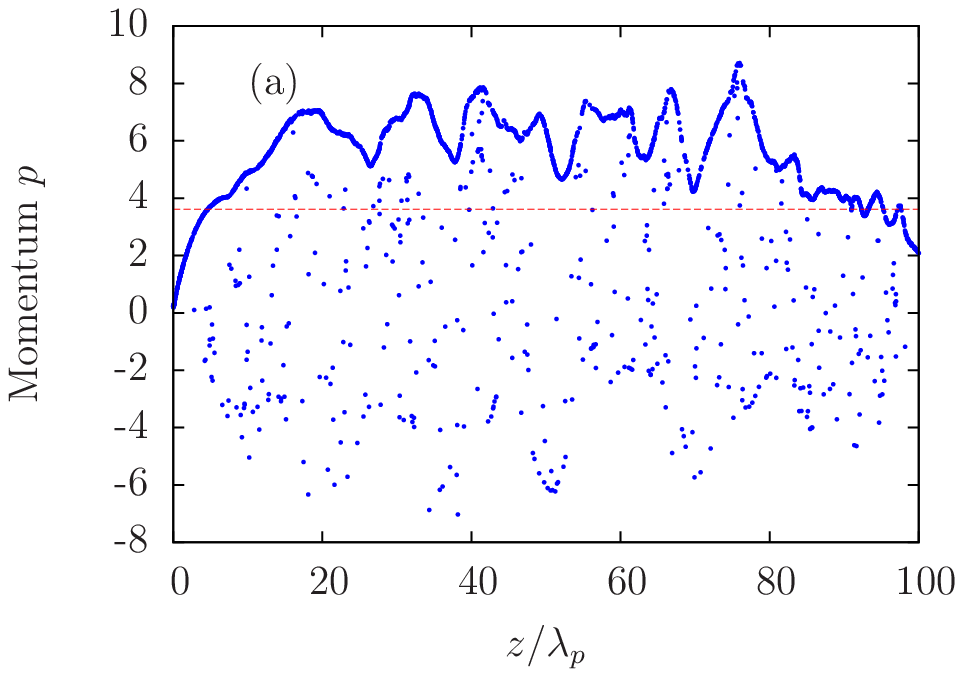}\\
  \plotone{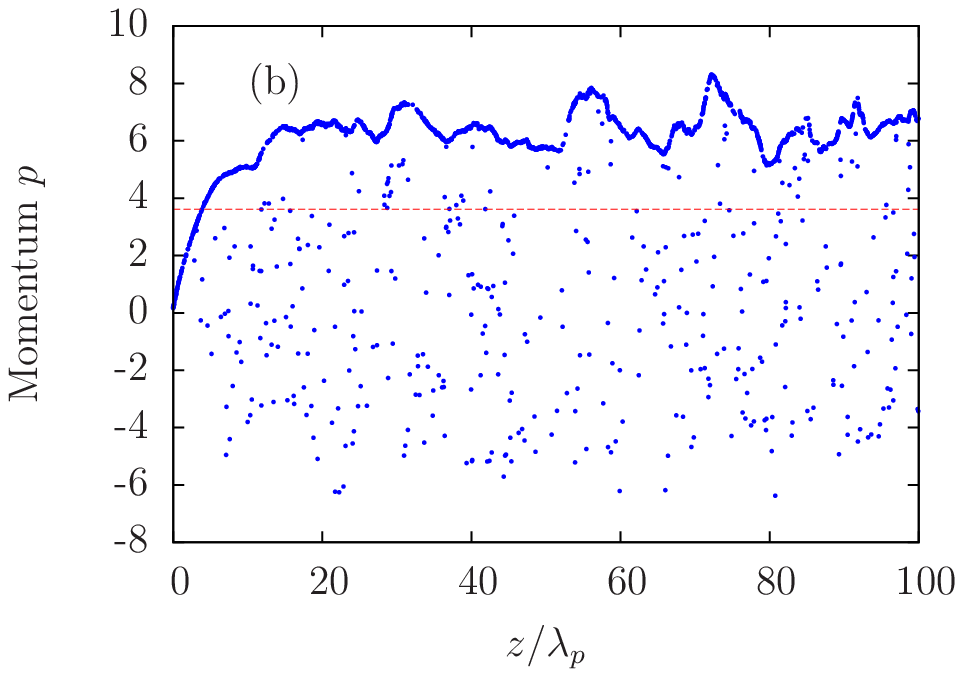}
\caption{Snapshot of 1000 randomly chosen particles in phase space for the flow 
with $\alpha=0.8$ and $\bar{\beta}_0=0.2$. 
Red dashed line shows the maximum momentum $p_{\max}$ for the steady cold 
solution with the same $\alpha=0.8$ and $\beta_0=0.2$. 
(a) Random snapshot for the simulation with box size $H = 100\lambda_p$. 
(b) Another random snapshot of a similar simulation with a larger computational box 
$H = 200\lambda_p$. 
 }
\label{fig:phasetraj}
\end{center}
\end{figure}

The moderate value of the hydrodynamic velocity does not, in principle, exclude
acceleration of a fraction of particles to much higher energies. We therefore also
studied the momentum distribution of particles in the flow.
Figure~\ref{fig:phasetraj}a shows a random snapshot of the particle distribution 
in the phase space for the flow with $\alpha=0.8$.
We randomly chose 1000 particles between $z=0$ and $z=100\lambda_p$
and the figure shows their locations in the two-dimensional phase space $(z,p)$.
The simulation demonstrates the following:

(1) There is no high-energy tail in the momentum distribution. 

(2) At each $z$, the momentum distribution has a pronounced narrow peak at 
$p_{\rm peak}$. Thus, a large fraction of particles form a cold stream;
this fraction is approximately equal to $\alpha$ (see below).
The momentum of the cold stream $p_{\rm peak}$ is above
(but comparable to) $p_{\max}$ predicted by the steady-state model. 

(3) There is a low-energy wing in the momentum distribution which extends to 
negative momenta. This broad component of the particle distribution has a 
hydrodynamic velocity close to zero; these particles are ``trapped'' and do not 
contribute much to the 
current density; however they make a significant contribution to charge density.
In our sample model, about 20\% of particles reside in the broad trapped 
component, 
and this fact has a simple explanation. From the point of view of the cold stream
dynamics, the broad component provides a background that offsets the effect 
of vacuum charge density $\rhoGJ$ by the fraction of 20\%. This fraction 
approximately equals to $1-\alpha$, so that the other particles 
(fraction $\approx \alpha$) may move in the cold 
stream with $v\approx c$ and carry $j_B$ without the mismatch in 
charge density that would generate strong $E_\parallel$. In essence, the broad 
component with backflowing particles allows the plasma to self-organize so that
the cold stream can keep $v\approx c$. This is in contrast to the steady-state 
solution in \Sect~2 where all particles formed a stream with a positive velocity 
$v\neq\alpha$, which must be periodically decelerated and accelerated.

(4) The cold stream momentum $p_{\rm peak}$
fluctuates in time (the corresponding curve in Figure~\ref{fig:phasetraj} moves in time).
However, the qualitative form of the 
phase-space distribution remains similar to that in Figure~\ref{fig:phasetraj}.

As seen in Figure~\ref{fig:phasetraj}a, the flow momentum $p_{\rm peak}$ decreases 
near the outer boundary of the computational box $z=H$.
This is an artifact of the boundary condition (free escape with no backflow), which 
suppresses the density of backflowing particles near the boundary. 
As a result, a modest negative electric
field is induced near the boundary, decreasing $p_{\rm peak}$ so that the flow carries
the required electric current $j_B$.
For comparison, Figure~\ref{fig:phasetraj}b shows a random snapshot of a similar model (in the same interval $0<z<100\lambda_p$) that has twice as large computational box, $H=200\lambda_p$.
As we increase $H$, the boundary effect moves away to larger $z$, affecting the flow 
properties only at $z\approx H$. The flow structure inside the box (away from the
boundary) does not depend on $H$.

\begin{figure}
\begin{center}
 \epsscale{1.15}
\plotone{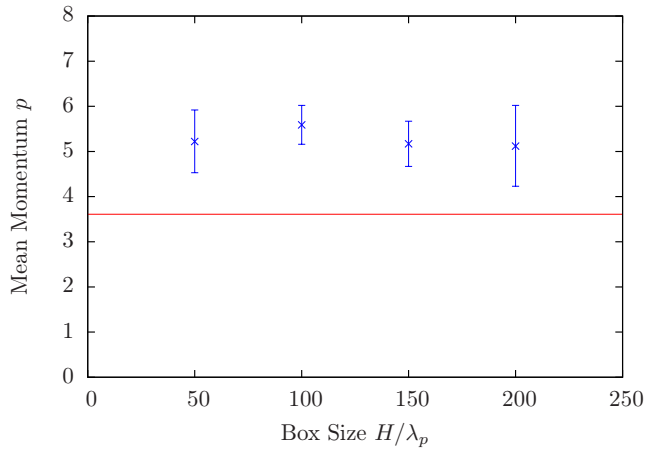}
\caption{Mean expectation and standard deviation for the fluctuating hydrodynamical 
momentum of the flow measured at the center of the computational box. Four 
simulations are shown, with box sizes $H/\lambda_p=50$, 100, 150, and 200;
all four simulations have the same parameter $\alpha=0.8$. Red line shows the 
maximum momentum predicted by the steady state model with $\alpha=0.8$.}
\label{fig:zetacompare}
\end{center}
\end{figure}

To check whether the flow momentum depends on the size of the computational 
box, we
ran several simulations with the same $\alpha = 0.8$ and different box sizes 
$H$. In each simulation, we measured the fluid momentum $\bar{p}$ in the center 
of the box (using a bin $\Delta z=2\lambda_p$) at time $t=100\omega_p^{-1}$.
The results are shown in Figure \ref{fig:zetacompare}.
There is no systematic variation in $\bar{p}$ with the box size; the small variations 
($\simlt 10$\%) are consistent with the fluctuations of $\bar{p}$ in time for each 
model.\footnote{
      Note that the average momentum $\bar{p}$ does not correspond to the average 
     velocity $\bar{v}$ shown in Figure~\ref{fig:meanvelocity}, in the sense that
     $\bar{p} \neq \bar{\beta}(1-\bar{\beta}^2)^{-1/2}$, because of the broad low-energy 
     tail of the distribution function. Compared with $\bar{v}$, the calculation of 
     $\bar{p}$ gives a higher weight to fast particles due to 
     the additional factor $\gamma$ in $p=\gamma\beta$.
     The average velocity remains close to $\alpha c$, and the averaged momentum 
     is larger than $\bar{\beta}(1-\bar{\beta}^2)^{-1/2}$. }

The polar-cap flows in pulsars extend through altitudes $z$ much larger than our 
box size $H$.
The fact that our results are independent of $H$ confirm the expected behavior 
--- the plasma keeps oscillating and particle acceleration is quenched 
everywhere as long as the flow satisfies the condition $0<\alpha<1$. 
We observe a quasi-uniform and quasi-steady behavior in the computational box
(apart from the initial acceleration region of length $\sim 10\lambda_p$).
In a realistic polar-cap flow, each segment of length $\sim 100\lambda_p$ should 
behave like our computational box.

We also ran simulations with varying $\alpha(z)$.
We ran models with $d\alpha/dz \sim 2\times 10^{-4}\lambda_p^{-1}$ (realistically,
$d\alpha/dz$ should be even smaller, $d\alpha/dz\simlt R^{-1}$, where $R$ is 
the star radius). The results are similar to the case of $\alpha=const$. 
The maximum momentum remains comparable to that given by \Eq~(\ref{eq:pmax}) 
as long as $0<\alpha<1$.


\section{Mixed ion flow and two stream instability}

If $j_B>0$ (which is equivalent to $\rhoGJ>0$ for $\alpha>0$), the 
charge-separated flow puled out from the polar cap is made of ions. 
Different ion species may end up in such a flow, and they will be 
accelerated to different velocities. 

The mixed ion flow shares many features with the identical-particle model 
studied in the previous sections. Steady state solutions 
can be obtained using the method described in \Sect~2.
Ions with different masses and charges move with different hydrodynamical 
momenta and co-exist in a common, periodic electrostatic potential. 
This steady solution
is prone to kinetic instability similar to that described in Sections~2 and 4.
There is, however, an important new feature: the ion streams with 
different hydrodynamical momenta are prone to two-stream instability.

To study the behavior of the mixed ion flow we slightly change the setup 
of our numerical simulation. Consider, e.g., 
a mixture of protons and helium nuclei (alpha-particles). 
The particle injection at $z=0$ now consists of two ions species; they have 
charges $e_1$ and $e_2=2e_1$, and masses $m_1$ and $m_2=4m_1$.
The two species are injected with equal rates $\dot{N}_1=\dot{N}_2$.
Then alpha-particles carry electric current $j_2=e_2\dot{N}_2$ that is
two times larger than the proton current $j_1=e_1\dot{N}_1$. Thus,
$j_2=(2/3)j_B$ and $j_1=(1/3)j_B$ are maintained at the boundary.

To define a characteristic plasma skin depth $\lambda_p$ we use 
\Eq~(\ref{eq:lambda}) where we replace $e,m,j_B$ by $(e_1,m_1,j_1)$ 
or, equivalently, by $(e_2,m_2,j_2)$ (note that $e_2j_2/m_2=e_1j_1/m_1$). 
The characteristic plasma frequency is defined by $\omega_p=c/\lambda_p$.

\begin{figure}
\epsscale{1.15}
\begin{center}
\plotone{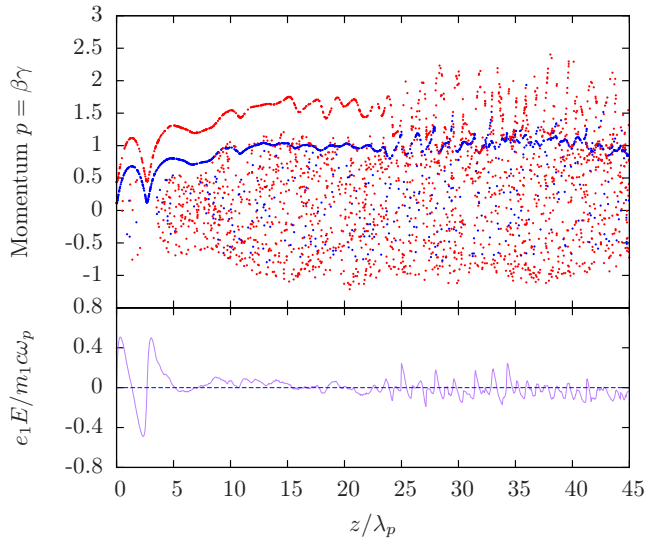}
\caption{Snapshot of the mixed ion flow with $\alpha=0.4$ at time 
$t = 10^3\omega^{-1}_p$.
Top panel shows the phase space distribution, where red 
dots represent protons and blue dots represent helium ions. 
Bottom panel  shows the electric field.
}
\label{fig:2stream}
\end{center}
\end{figure}

Figure~7 shows a snapshot of the phase-space distribution of ions long after 
the beginning of the simulation. In this sample model $\alpha=0.4$. The
modest value of $\alpha$ (not close to unity) implies modest Lorentz factors 
of particles and the fast development of instabilities. 
The flow exhibits the following features:

\noindent
(1) One period of the steady state solution is reproduced near the injection 
boundary $z=0$. The period $z_0\approx 3\lambda_p$ agrees with the result 
from numerical integration of the corresponding steady state model.
(This feature is stable in our sample model 
because we chose a high injection velocity $v_0\approx 0.4c$.)

\noindent
(2) At larger $z$ the periodic flow becomes unstable and develops into a 
configuration similar to that in Figure~5, except that now we have {\it two}
cold variable streams. Besides the cold streams, there is a broad distribution 
of ions with smaller momenta and a negligible hydrodynamic velocity.
The origin of this broad component of trapped particles was discussed in 
\Sect~4.2 and plays here a similar role --- it is self-organized 
so that the streams may move with a relativistic speed without 
a mismatch in charge density.

\noindent
(3) Further from the boundary (at $z>20\lambda_p$), the two streams
develop a two-stream instability. The growth rate of the instability may be 
estimated using an idealized model of two cold fluids with densities $n_1$,
$n_2$ and velocities $v_1,v_2$. It is straightforward to derive the dispersion 
relation for Langmuir modes with frequency $\omega$ and wave-vector $k$
(e.g. Melrose 1986); it gives,
\begin{equation}
\label{eq:disp}
    1-\frac{\omega_1^{2}}{\gamma_1^3(\omega - kv_1)^2} - \frac{\omega_2^{2}}{\gamma_2^3(\omega - kv_2)^2} = 0,
\end{equation}
where $\omega_1^2=4\pi n_1 e_1^2/m_1$ and $\omega_2^2=4\pi n_2 e_2^2/m_2$.
Using $\omega_1\approx\omega_2\approx\omega_p$ and the characteristic values
of velocities $v_1,v_2$ from our simulation, we find from \Eq~(\ref{eq:disp}) that 
the most unstable modes have $\omega$ comparable to $\omega_p$ and their
growth rate is $\Gamma\sim 0.2\omega_p$. In the simulation we observe a 
slightly smaller $\Gamma$. The distance over which the
Langmuir waves are amplified is roughly  $10\lambda_p$.
As a result of the instability, the two streams are smeared out at large $z$, 
in particular the stream of lighter ions. 
No significant particle acceleration is seen in the simulation.


\section{Discussion}

We have presented detailed one-dimensional time-dependent simulations 
of the plasma flow extracted from the polar caps of neutron stars.
The simulations provide a fully kinetic description of the flow,
with self-consistent electric field and particle distribution function.
In this paper, we focused on the regime $0<\alpha<1$, where $\alpha$ is the main 
parameter of the flow defined by \Eq(\ref{eq:alpha}).
In agreement with the estimates of B08, we find that 
the particles are accelerated to Lorentz factors, 
\beq
\label{eq:gam2}
   \gamma\approx \frac{1+\alpha^2}{1-\alpha^2},
 \eeq
and are not capable of igniting pair creation.
In this sense, flows with $0<\alpha<1$ are ``dead.''
They are sustained by a modest voltage, oscillating in space and time.
Although the simulation is limited to regions close to the pulsar surface,
the result does not depend on the simulation box size,
and hence should describe the entire polar cap flow,
as long as $\alpha$ remains between $0$ and $1$.
The parameter $\alpha$ is expected to vary along the 
magnetic field lines, on a scale comparable to the stellar radius;
we have verified that this variation does not change the oscillating behavior
of the flow (see also B08).

The simulations show how a kinetic instability develops and disrupts the ideal 
periodic structure found in the analytical models of the dead zone;
the instability mechanism is described in \Sects~2 and 4.
We find that the momentum distribution function has two distinct parts --- a 
variable ``cold stream'' and a broad wing at low momenta, which includes particles 
flowing backward to the polar cap. The fraction of particles in the cold stream 
is approximately equal to $\alpha$; the remaining fraction $1-\alpha$ forms
the broad component. Even though the flow is turbulent, it shows 
no signs of particle acceleration to energies higher than that of the cold stream. 

The value of parameter $\alpha$ depends 
on the location and geometry of the polar cap.
A simplest magnetospheric configuration is that of a centered dipole. Then 
the parameter $\alpha$ depends on the angle between the magnetic and spin axes, 
$\xi$; besides, it varies across the polar cap. For nearly
aligned rotators ($\xi\approx 0$), 
$0<\alpha<1$ in the central part of the polar cap and $\alpha<0$
in a ring-shaped zone near the edge of the polar cap 
(Timokhin 2006; Parfrey et al. 2012).
In this case, the dead zone occupies the 
central part of the polar cap, and $e^\pm$ discharge must be confined to the ring,
matching the phenomenological ``hollow cone'' model of pulsar emission.
In contrast, the polar cap of an orthogonal rotator ($\xi\approx \pi/2$) has 
$|\alpha|\gg 1$, which enables $e^\pm$ discharge for the entire polar cap. 
At arbitrary misalignment $0<\xi<\pi/2$,
the values of $\alpha$  are provided by global three-dimensional simulations 
of the magnetospheric structure (e.g. Spitkovsky 2006) and should play a key 
role for the geometry of the radio beam.

We presented our results using plasma skin depth $\lambda_p$ as a unit of 
length and particle rest-mass $mc^2$ as a unit of energy. In this form, the results 
do not depend on the charge or mass of the particles extracted from the polar cap,
as long as the flow is made of identical particles. 
In particular, \Eq~(\ref{eq:gam2}) is valid for both electron flow 
($\rhoGJ<0$) and ion flow ($\rhoGJ>0$), and the phase-space 
distribution shown in Figure~5 describes both cases.
Note that the accelerating voltage is proportional to the particle mass; 
voltage implied by \Eq~(\ref{eq:gam2}) is different for ions and electrons by 
the factor of $m_i/m_e\sim 2\times 10^3$.
The relatively high voltage in the ion flow, 
$e\Phi\approx m_ic^2 (1+\alpha^2)/(1-\alpha^2)$
is still hardly sufficient to ignite $e^\pm$ pair discharge by a seed electron or positron.

The identical-particle model may not hold for an ion flow; in this case, 
new effects may enter the problem.
Firstly, heavy ions pulled out from the polar cap may not be completely ionized 
and begin to lose electrons as they are accelerated and interact with the X-rays 
above the stellar surface; this process effectively creates new charges, 
reminiscent of pair creation (e.g. Jones 2012). Secondly, the ion flow may be a 
mixture of different nuclei which will be accelerated to different Lorentz factors. 
The mixed ion flow is prone to two-stream instability, possibly leading to formation
of plasma clumps and generation of coherent radio emission.
In our simulations, we observe the expected two-stream instability,
however do not observe significant structure (clumps) in the turbulent flow.
This may change in three-dimensional simulations.
The frequency of excited waves (comparable to the ion plasma frequency) 
is in the radio band, and coherent emission from clumps could create bright 
coherent emission. It remains to be seen whether this mechanism
can contribute to the pulsar emission. If it does, it would create an additional 
component of the radio pulse. In the case of approximately aligned rotator, 
the additional component would be generated in the central region of the polar 
cap, leading to a ``hollow cone + core'' structure of the radio pulse. 

The charge-separated model of the dead zone can be modified to include possible 
backflowing particles from distant parts of the open field-line bundle 
(e.g. from a pair-producing outer gap). These particles can contribute to the current 
density and also serve as an additional background charge density, which may be 
modeled as a contribution to the effective ``vacuum'' charge density  $-\rhoGJ$. 
This would change the effective $\alpha$ (Lyubarsky 1992; B08), most likely 
reducing it.

An outer gap is expected to form in a charge-separated flow near the null surface 
$\vec{B}\cdot\vec{\Omega}=0$ (Cheng et al. 1986). On a given field line, 
the outer gap will be screened if it is loaded by multiple $e^\pm$ pairs produced by 
the discharge at the polar cap. Thus, the
suppression of $e^\pm$ discharge near the field-line footpoint is 
an essential condition for the existence of an outer-gap accelerator. 
Therefore, one can expect an outer gap to form on 
field lines with footpoints in the dead zone.

We did not simulate in this paper flows with $\alpha>1$ or $\alpha<0$; in these 
cases particles must be strongly accelerated. This regime leads to $e^\pm$ 
discharge that must be unsteady, with a significant intermittent backflow (B08). 
A model for oscillating discharge may be constructed in hydrodynamical 
approximation (Levinson et al. 2005), however a fully kinetic description is essential, 
as demonstrated by our results for the dead zone, where a significant fraction of 
particles are trapped and form a broad wing at low momenta in the distribution 
function. The discharge simulation can be done using our setup of a fixed current 
$j=j_B$ at the stellar surface (BT07) and incorporating pair creation.
We defer the simulations with $\alpha>1$ and $\alpha<0$ to a future work. 

When this work was completed, the preprint by Timokhin \& Arons (2012) came out. 
They present simulations of charge separated flows, using a similar method,
and the results agree with our results for $0<\alpha<1$. They also consider flows 
with $\alpha<0$ and $\alpha>1$, and find a strong unsteady $e^\pm$ discharge 
confirming the analysis in B08.

\acknowledgements
This work was supported by NASA NNX10AI72G.



\end{document}